\documentclass[useAMS,usenatbib]{mn2e}
\usepackage{graphicx}
\usepackage{amssymb}

\def\aj{AJ}
\def\apj{ApJ}
\def\apjl{ApJL}
\def\aap{A\&A}

\def\mnras{MNRAS}
\def\nat{NATURE}

\def\procspie{Proceedings of the SPIE}

\def\pasp{Publications of the Astronomical Society of the Pacific}
\def\araa{Annual Review of Astronomy and Astrophysics}

\def\iaucirc{International Astronomical Union Circulars}

\title{iPTF13beo: The Double-Peaked Light Curve of a Type Ibn Supernova Discovered Shortly after Explosion}
\author[Gorbikov et al.]
  {Evgeny Gorbikov$^1$, Avishay Gal-Yam$^1$, Eran O. Ofek$^1$, Paul M. Vreeswijk$^1$, \and Peter E. Nugent$^{2,3}$, Nicolas Chotard$^4$, Shrinivas R. Kulkarni$^5$, Yi Cao$^5$,\and 
  Annalisa De Cia$^1$, Ofer Yaron$^1$, David Tal$^1$, Iair Arcavi$^1$, Mansi M. Kasliwal$^6$,\and 
  S. Bradley Cenko$^{7,8}$, Mark Sullivan$^9$, Juncheng Chen$^{10}$
\\
  $^1$Benoziyo Center for Astrophysics, Faculty of Physics, The Weizmann Institute of Science, Rehovot 76100, Israel \\
  $^2$Department of Astronomy, University of California Berkeley, B-20 Hearst Field Annex \# 3411, Berkeley, CA 94720-3411, USA \\
  $^3$Computational Cosmology Center, Computational Research Division, Lawrence Berkeley National Laboratory,\\ ~~~1 Cyclotron Road MS 50B-4206, Berkeley, CA 94720, USA \\
  $^4$Universit\'e de Lyon, F-69622, Lyon, France; Universit\'e de Lyon 1, Villeurbanne ;\\ ~~~CNRS/IN2P3, Institut de Physique Nucl\'eaire de Lyon. \\
  $^5$Astronomy Department, California Institute of Technology, 1200 E. California Boulevard, Pasadena, CA 91125, USA \\
  $^6$The Observatories, Carnegie Institution for Science, 813 Santa Barbara Street, Pasadena, CA 91101, USA \\
  $^7$Astrophysics Science Division, NASA Goddard Space Flight Center, Mail Code 661, Greenbelt, MD 20771, USA \\
  $^8$Joint Space Science Institute, University of Maryland, College Park, Maryland 20742, USA \\
  $^9$School of Physics \& Astronomy, University of Southampton, Highfield, Southampton SO17 1BJ, U.K. \\
  $^{10}$Tsinghua Center for Astrophysics, Tsinghua University, Beijing 100084, China
}

\begin{document}
\maketitle

\begin{abstract}
We present optical photometric and spectroscopic observations of the Type Ibn (SN 2006jc-like) supernova iPTF13beo. Detected by the intermediate Palomar Transient Factory $\sim$3 hours after the estimated first light, iPTF13beo is the youngest and the most distant ($\sim$430 Mpc) Type Ibn event ever observed.
The iPTF13beo light curve is consistent with light curves of other Type Ibn SNe and with light curves of fast Type Ic events, but with a slightly faster rise-time of two days. In addition, the iPTF13beo $R$-band light curve exhibits a double-peak structure separated by $\sim$9 days, not observed before in any Type Ibn SN. A low-resolution spectrum taken during the iPTF13beo rising stage is featureless, while a late-time spectrum obtained during the declining stage exhibits narrow and intermediate-width He\,I and Si\,II features with FWHM $\approx$ 2000--5000\,km\,s$^{-1}$ and is remarkably similar to the prototypical SN Ibn 2006jc spectrum. We suggest that our observations support a model of a massive star exploding in a dense He-rich circumstellar medium (CSM). A shock breakout in a CSM model requires an eruption releasing a total mass of $\sim$0.1\,M$\odot$ over a time scale of couple of weeks prior to the SN explosion.
\end{abstract}

\begin{keywords}
 (stars:) supernovae: general, (stars:) supernovae: individual: iPTF13beo, stars: Wolf-Rayet, stars: winds, outflows
\end{keywords}

\section{Introduction}

Among hydrogen-poor supernovae (Type I SNe) a new and rare subtype, SNe Ibn, has recently emerged with the discovery of SN 1999cq (\citealt{MAT00,PAS08I}). Type Ibn SNe are characterized by the presence of narrow or intermediate-width ($\sim$2000--3000\,km\,s$^{-1}$) He I emission lines, which are attributed to the SN ejecta interacting with He-rich CSM (\citealt{PAS08I}). Several lines of evidence suggest that Type Ibn SN progenitors are massive stars characterized by high mass-loss rates and stripped He-rich envelopes, such as Wolf-Rayet stars (\citealt{FOL07, PAS07, PAS08I,SAN13}). A luminous outburst coincident with the location of Type Ibn SN 2006jc, occurring two years before the SN explosion, was reported by \cite{NAK06}. Such an outburst is considered to be a likely source for the He-rich dense CSM around some Type Ibn SNe (\citealt{PAS07,FOL07}).

Eight Type Ibn SNe have been identified to date. The most well studied among them is SN 2006jc (\citealt{NAK06,PAS07,PAS08I,FOL07,IMM08,SMI08,ANU09,OFE13a}). The other SN 2006jc-like events are: SN 1999cq (\citealt{MAT00,PAS08I}), SN 2000er (\citealt{PAS08I}), SN 2002ao (\citealt{MAR02,FIL02,FOL07,PAS08I}), PS1-12sk (\citealt{SAN13}) and iPTF13beo described here. Two more peculiar SNe, SN 2005la (\citealt{PAS08II}) and SN 2011hw (\citealt{SMI12}), are assumed to be of a transitional case between Type Ibn and Type IIn, since they show strong and narrow or intermediate-width lines of both hydrogen and helium.

No Type Ibn SNe were discovered well before $R$-band brightness maximum so far, therefore their early light curve evolution was not known. Probably the youngest one is SN 1999cq (\citealt{MAT00,PAS08I}) discovered $\sim$1 day before $R$-band maximum. Unfortunately, the first SN 1999cq spectrum was obtained $\sim$19 days after the $R$-band maximum and the opportunity for studying Type Ibn SN early spectra was missed. The earliest spectral observations of Type Ibn SNe were obtained for SN 2000er (\citealt{PAS08I}) with the first spectrum obtained $\sim$5 days after the estimated explosion date. 

In this Paper, we present photometric and spectroscopic observations of the Type Ibn event iPTF13beo. Discovered $\sim$2 days before the $R$-band maximal brightness, iPTF13beo provides a unique opportunity to study the early light curve and pre-maximum spectral evolution. At $z=0.091$ or 431 Mpc\footnote{Adopting a standard $\Lambda$CDM cosmology and a Hubble constant $H_0=67.8$\,km\,s$^{-1}$\,Mpc$^{-1}$ (\citealt{PLA13}).} iPTF13beo is twice as distant as the most distant Type Ibn SN discovered so far, PS1-12sk ($z=0.054$, \citealt{SAN13}). 

\section{Observations and Analysis}
\subsection{Discovery}

\begin{figure*}
\begin{center}
\begin{tabular}{cc}
\includegraphics[angle=0,width=0.4767\textwidth]{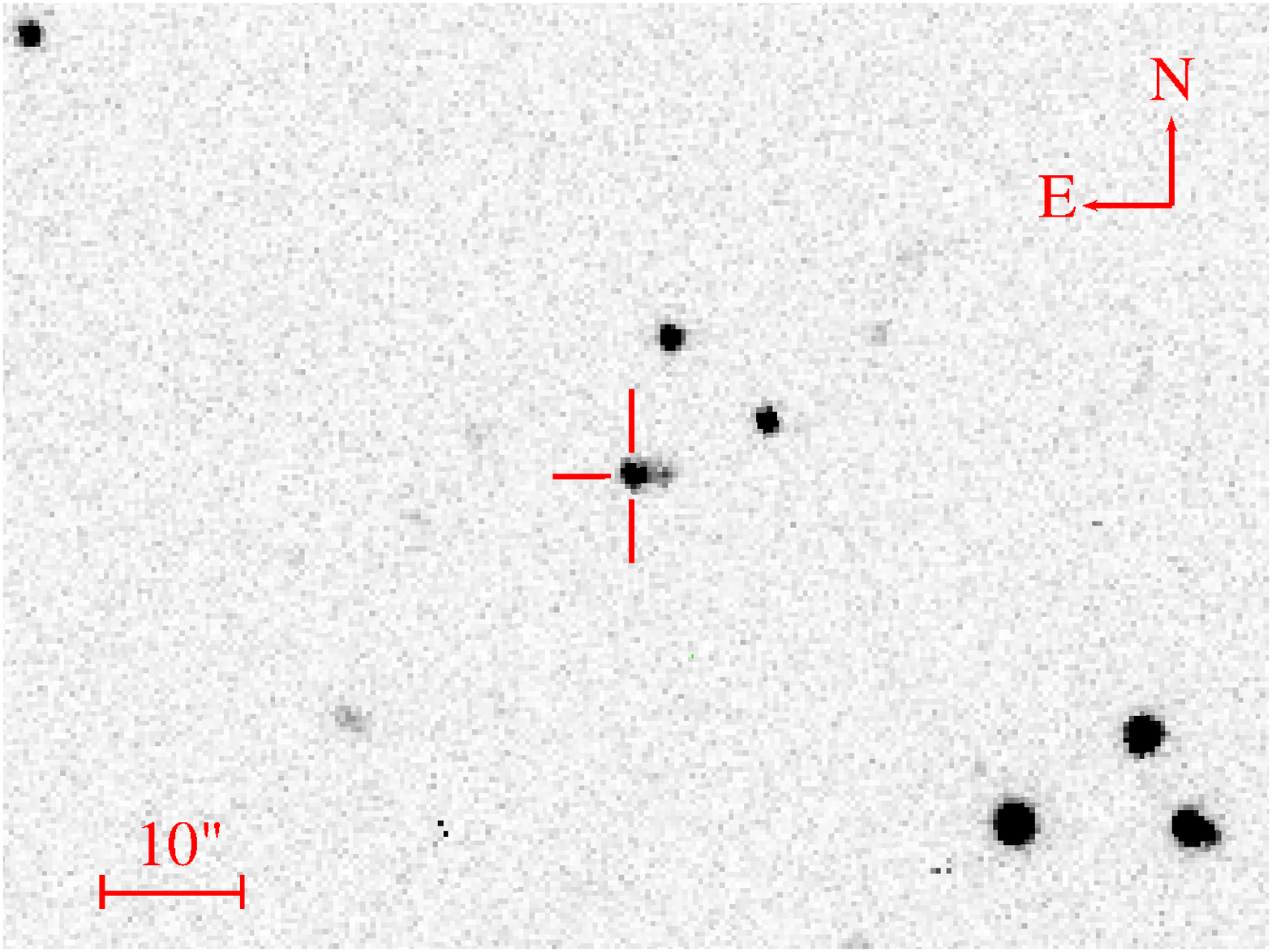} &
\includegraphics[angle=0,width=0.4766\textwidth]{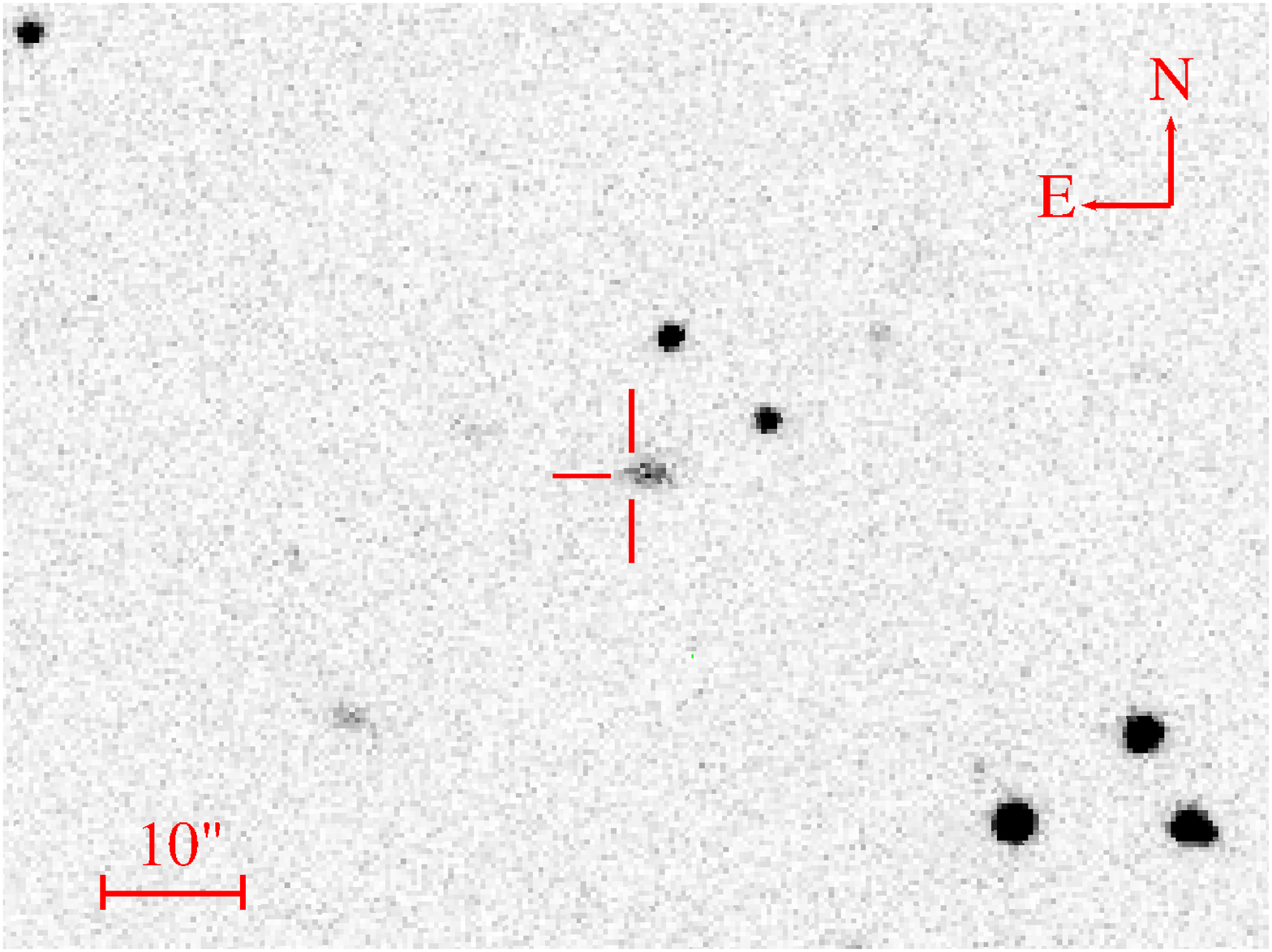} \\
$a$ & $b$ \\
\end{tabular}
 \end{center}
\caption{$a$: A 120-sec $R$-band image of iPTF13beo close to maximum, taken on 2013 May 21.474 using the P60 telescope (\citealt{CEN06}). $b$: A late-time 120-sec $R$-band image with a negligible iPTF13beo flux taken on 2013 July 7.340 using the P60 telescope. The SN location is marked by the reticle.\label{iPTF2013beo}}
\end{figure*}

iPTF13beo was discovered on UTC 2013 May 19.349 (JD 2456431.849)\footnote{All dates are specified in the UTC system.} by the intermediate Palomar Transient Factory (iPTF; \citealt{LAW09,RAU09,KUL13}) at $R=20.9\pm0.2$\,mag. The source was detected at $\alpha=16^h12^m26.63^s$, $\delta=+14^\circ19'18.0''$ J2000, $\sim$1.4 arcsec apart from the center of the faint ($u\approx20.9$\,mag, $r\approx19.5$\,mag) galaxy SDSS J161226.53+141917.8. This faint host galaxy was never classified morphologically before, however it can be classified as a blue late-type galaxy using its colour and absolute magnitude, following the \cite{BAL04} definition. Nothing was detected at the SN location two days before, on May 17.348, down to $R\approx21.1$\,mag (5$\sigma$ detection threshold). Figure \ref{iPTF2013beo} shows the SN image near peak $R$-band brightness and a reference image without SN light taken $\sim$1.5 months later.

\subsection{Photometry}

\begin{figure*}
\begin{center}
 \includegraphics[angle=0,width=0.875\textwidth]{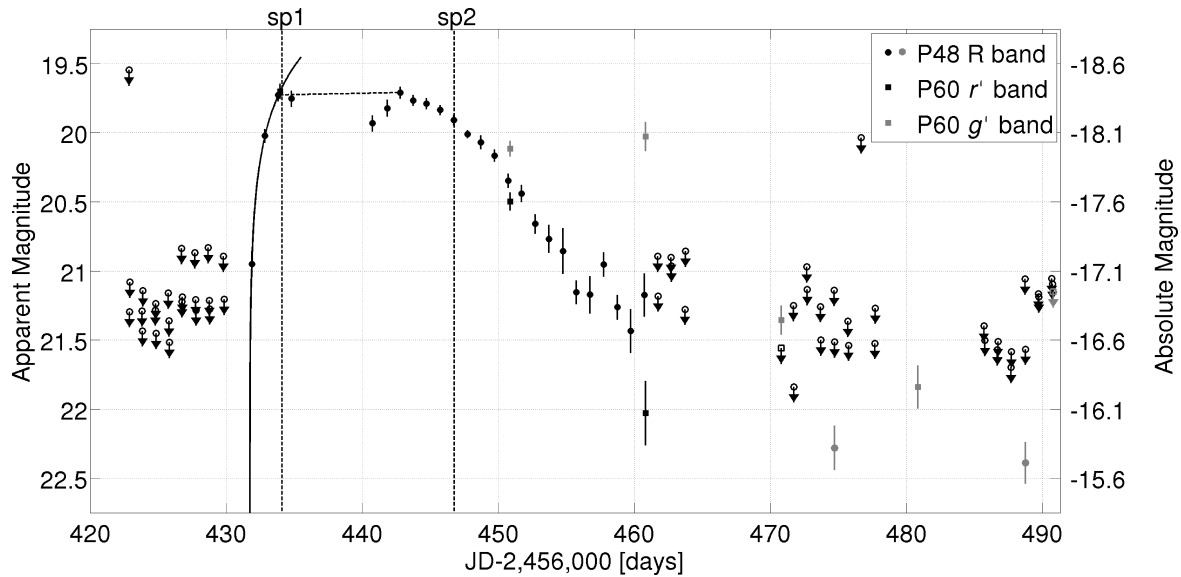} 
 \end{center}
\caption{iPTF13beo light curve. The filled black dots show the P48 R-band nightly-binned average SN magnitudes. The filled gray dots show the P48 R-band late-time SN magnitudes obtained by the image co-adding. The filled black and gray squares show the P60 $g'$-band and $r'$-band magnitudes correspondingly. The empty symbols show the upper limits for the SN non-detections. The black vertical dashed lines marked as `sp1' and `sp2' show the JDs of the SN spectral observations. The solid black line shows the power law model fitted to the light curve rising part. The black dashed nearly horizontal line is shown to demonstrate the double-peak model statistical significance. \label{iPTF2013beo_lc}}
\end{figure*}

Optical observations were performed mainly using the 48-inch Samuel Oschin Schmidt Telescope at Palomar Observatory (P48) equipped with the CFH12K 96 Mpx mosaic camera (\citealt{RAH08}). The iPTF13beo field was monitored at least twice per night. The observations were performed with a Mould R filter (\citealt{OFE12}) and the photometry was obtained from image subtraction using the Sullivan pipeline via forced PSF fitting, calibrated to SDSS stars in the field. Several iPTF13beo observations were performed using the robotic 60-inch telescope at Palomar Observatory (P60; \citealt{CEN06}) in the $g'$ and $r'$ bands and were also calibrated to SDSS stars. Figure \ref{iPTF2013beo_lc} shows the iPTF13beo light curve, as well as non-detection upper limits. 

We estimate the explosion time by fitting a simple power law model $f(t)=C(t-t_0)^\beta$ to the light curve in flux units in a similar manner to \cite{NUG11}. Here $t_0$ is the explosion time. The model is fit only to the rising part of the light curve, i.e., to the P48 R-band detections on May 19, 20 and 21 and the P60 $r'$-band detection on May 21. In the simplest case of an expanding fireball, the exponent is kept constant, $\beta=2$, and the best fit yields $t_0=2,456,431.32\pm0.22$ ($\chi^2/dof\cong6.6/2$). Allowing the exponent to vary from 2, the best fit is estimated to be $\beta=1.08\pm0.27$, $t_0=2,456,431.73\pm0.10$ ($\chi^2/dof\cong0.011/1$), yielding an explosion date of UTC 2013 May 19.23 $\pm$ 0.10. The latter fit is shown in Figure \ref{iPTF2013beo_lc}. The first detection of the SN by our pipeline was $\sim$3 hours after the estimated explosion time. However, the SN remained unnoticed by observers till the next night, May 20, when it was flagged by the duty astronomer as a rapidly 
rising ($\sim$1\,mag per night) SN candidate at $R\approx20$\,mag.   

The iPTF13beo light curve is characterized by a very fast ($\sim$2 days) rise, reaching a peak apparent brightness $R=19.72\pm0.04$\,mag (absolute: $-18.36$\,mag) on May 21. On May 22 the SN brightness ($R=19.75\pm0.06$\,mag) did not change significantly. Unfortunately, during the next five nights, from May 23 to May 27, iPTF13beo was not monitored. On May 28, nine days after the estimated explosion time, the SN began rising again from $R=19.93\pm0.06$\,mag, reaching its second peak $R=19.71\pm0.04$\,mag (absolute: $-18.37$\,mag) on May 30, $\sim$11 days after the estimated explosion date and $\sim$9 days after the primary peak. After that, the SN constantly declined at an average rate of $\sim$0.1\,mag per day for 18 days, until it faded below our detection limit of $R\approx21$\,mag (5$\sigma$) on June 18, being visible for a total of 29 nights. Two more late-time data points were obtained on July 1 and July 15 using image co-adding.

Our assumption of a second peak in the iPTF13beo light curve is based on two data points, those on May 28 and May 29. Without them the light curve could be fit with a single-peak model. To demonstrate the statistical significance of the second peak, we draw a straight line between the points at May 21 and May 30, ignoring the points taken on May 22, May 28 and May 29. This line represents a simple single-peak model and is shown in the inset of Figure \ref{iPTF2013beo_lc}. The May 28 point falls $\sim$3.7 times its photometric error below the line, while the May 29 point is $\sim$1.8 times its photometric error below the line. Multiplying the one-sided Gaussian cumulative probabilities we obtain the probability of 0.0004\% to get two points below this line. The May 22 point was not taken into account, since it does not show any significant change in brightness relative to the previous night. We conclude that the single-peak model can be ruled out in favor of the double-peak model. 

This double-peak light curve structure was not observed in any previous Type Ibn SN. However, none were observed during the rising phase, as mentioned above. Therefore, this double-peak structure might be a characteristic feature of Type Ibn light curves. We interpret the two peaks of the light curve as follows: the first peak is explained by the SN shock breakout in a dense CSM (e.g., \citealt{OFE10}), while the second peak is explained by the SN regular radioactive decay modified by the interaction of the SN ejecta with a CSM. This explanation is also supported by spectroscopic observations, as will be shown later.   

\subsection{Light Curve Comparison}

\begin{figure*}
\begin{center}
\includegraphics[angle=0,width=0.843\textwidth]{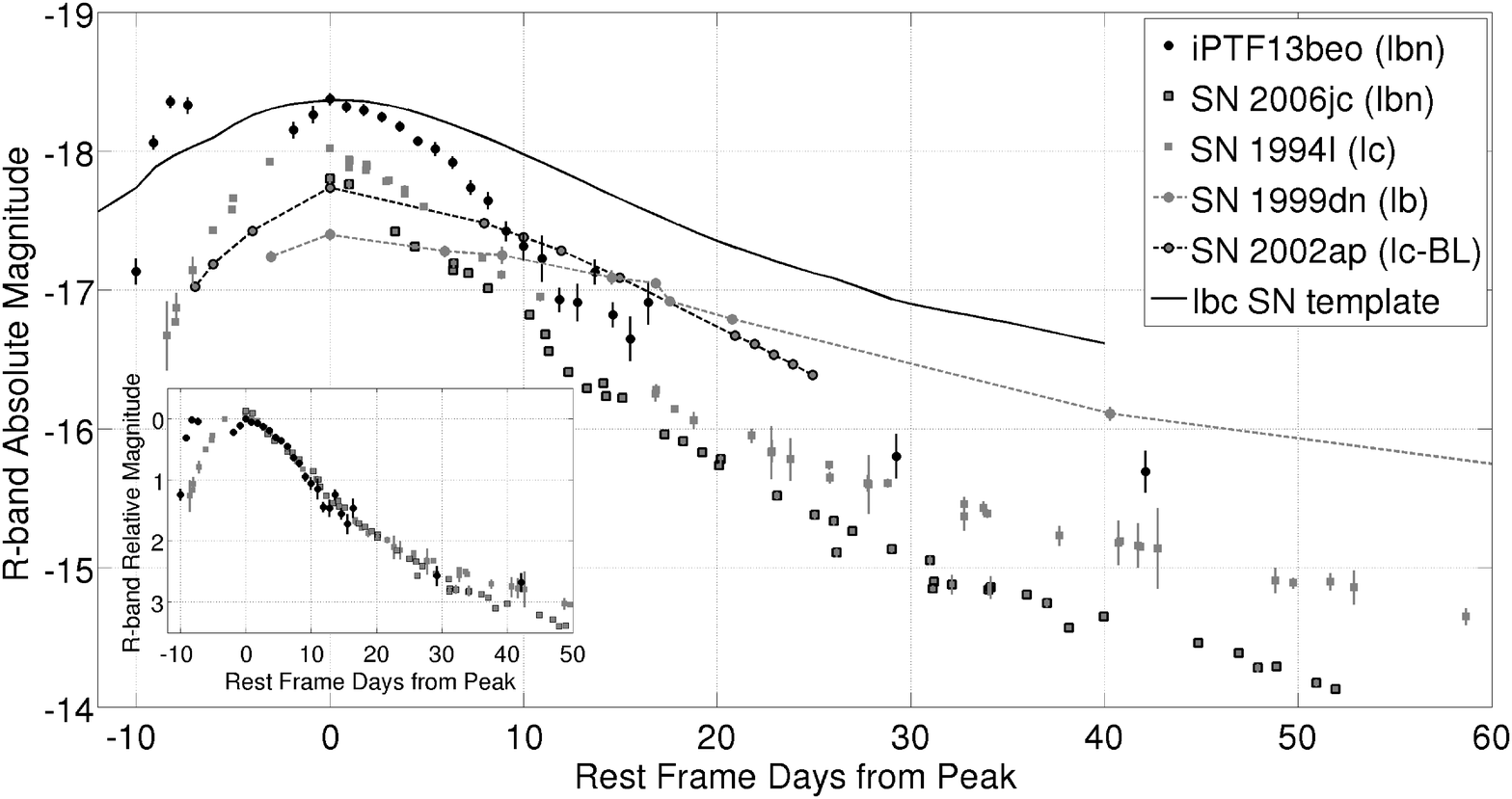}  
 \end{center}
\caption{Comparison between the iPTF13beo light curve and those of Type Ib, Ic and Ibn SNe (absolute magnitudes). SN 2006jc (Ibn) light curve is taken from \protect\cite{PAS07,PAS08I} and \protect\cite{ANU09}, SN 1994I (Ic) data - from \protect\cite{SN94I}, SN 1999dn (Ib) - from \protect\cite{SN99dn}, SN 2002ap (Ic-BL) - from \protect\cite{GAL02}, Type Ibc template - from \protect\cite{DRO11}. The horizontal axis shows the rest frame time from the $R$-band peak, in the case of iPTF13beo - the rest frame time from the second peak. 
The inset shows the iPTF13beo post-peak light curve fitted with SN 1994I and SN 2006jc light curves. \label{lc_comp}}
\end{figure*}

Figure \ref{lc_comp} shows the iPTF13beo light curve in absolute magnitudes and the light curves of other SN types for comparison. From visual inspection we conclude that iPTF13beo rises and declines more rapidly that a typical SN Ibc. We also note that the iPTF13beo post-peak light curve is similar to those of SN 2006jc (Ibn) and SN 1994I (fast Ic). The similarity of Type Ibn post-peak light curves to those of fast Type Ic events was noted already by \cite{MAT00} using the observations of SN 1999cq. Later it was also demonstrated by \cite{FOL07} for the best studied Type Ibn SN 2006jc. 

To quantify this similarity we fitted the iPTF13beo post-peak light curve with those of SN 2006jc and SN 1994I using only time and magnitude shifts as parameters. The inset of Figure \ref{lc_comp} shows the best fit of the iPTF13beo post-peak photometric data to the SN 2006jc light curve ($\chi^2/dof\cong52.9/18$) and to the SN 1994I light curve ($\chi^2/dof\cong30.2/18$). We conclude that the iPTF13beo post-peak light curve is consistent with Type Ibn and with fast Type Ic light curves.

The iPTF13beo rising-phase light curve was also fitted with the SN 1994I light curve yielding $\chi^2/dof\cong31.6/2$. The reason for this inconsistency is the fact that iPTF13beo rises more rapidly than even this fast SN Ic. 

\subsection{Spectroscopy}

\begin{figure*}
\begin{center}
\includegraphics[angle=0,width=0.843\textwidth]{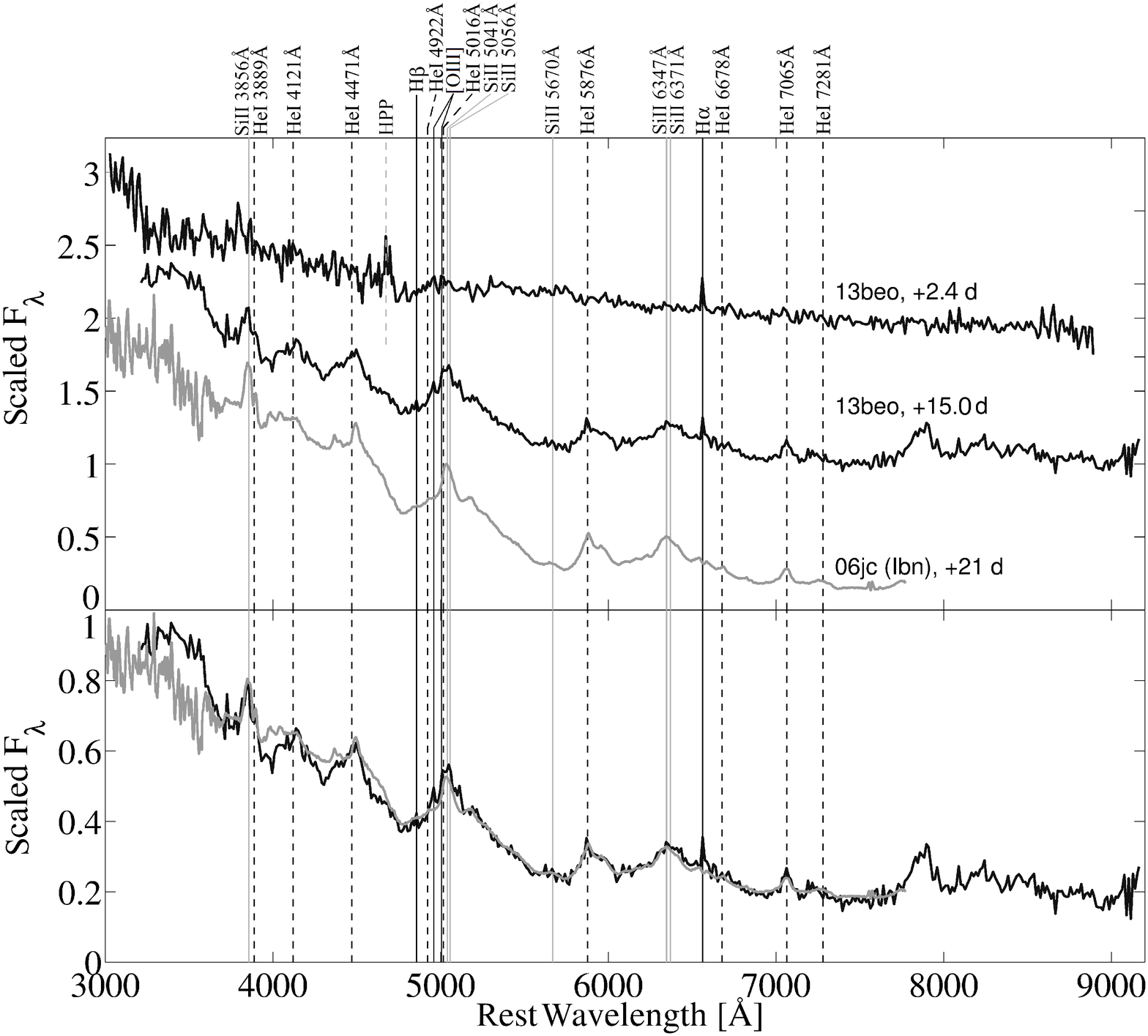}  
 \end{center}
\caption{$Top$ $panel$: The spectra of iPTF13beo (black) at phases +2.4 days and +15.0 days compared to the spectrum of SN 2006jc (Ibn, gray) at phase +21 days. The SN 2006jc spectrum is from \protect\cite{BUF09}. The iPTF13beo phases are defined relative to the estimated explosion time, while the SN 2006jc phase is determined relative to the first detection. The locations of H$\alpha$, H$\beta$ and [O\,III] lines are marked by solid black lines, the locations of He\,I lines are denoted by dashed black lines and the locations of Si\,II lines are shown with solid gray lines. The dashed gray line shows the Half-Power Point between two SNIFS channels (\citealt{BUT13}) redshifted to the iPTF13beo host redshift.
$Bottom$ $panel$: The late (+15.0 days) iPTF13beo spectrum (black) overplotted with the SN 2006jc (+21 days) spectrum (gray).
\label{spec_general}}
\end{figure*}

\begin{figure}
\begin{center}
\includegraphics[angle=0,width=0.476\textwidth]{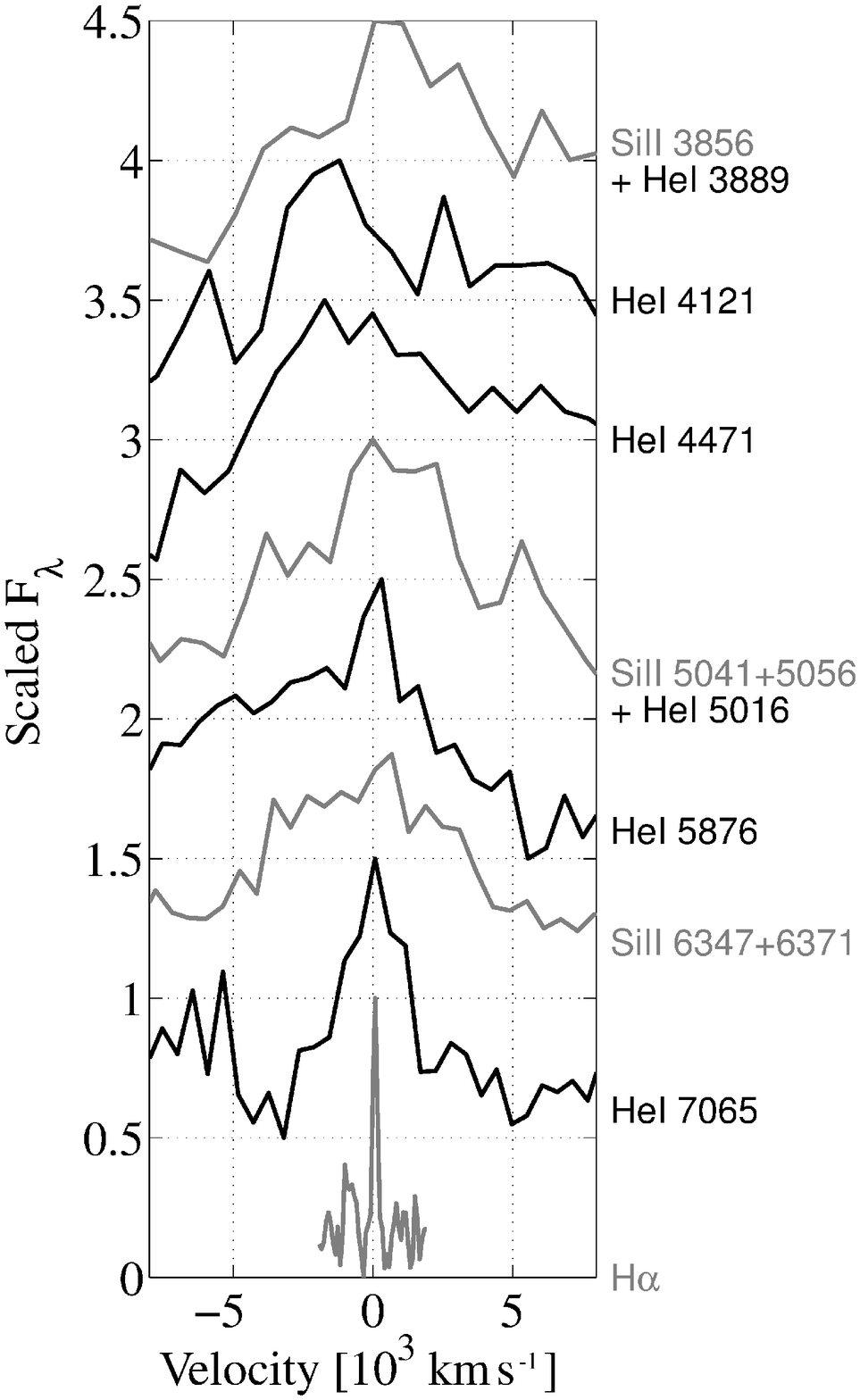}  
 \end{center}
\caption{Low-resolution velocity profiles of iPTF13beo lines from the late spectrum (phase: +15.6 days). See the legend for the wavelengths.
\label{spec_lines}}
\end{figure}

Low-resolution spectroscopic follow-up of iPTF13beo was performed on the dates specified graphically in Figure \ref{iPTF2013beo_lc}. Two spectra were taken. The first one, referred to here as the `early spectrum', was taken on 2013 May 21.594, $\sim$2.2 days after the first SN detection, $\sim$2.4 days after the estimated SN explosion date (phase: +2.4 days\footnote{All phases here are estimated relative to the approximate explosion date.}) and $\sim$0.3 days before the primary peak. The spectrum was obtained using the Supernova Integral Field Spectrogaph (SNIFS, \citealt{LAN04,BUT13}) mounted on the University of Hawaii 2.2-meter telescope (88-inch, UH88).

The second SN spectrum, referred to here as the `late spectrum', was taken on 2013 June 3.277, $\sim$14.9 days after the first detection, $\sim$15.0 days after the estimated SN explosion (phase: +15.0 days) and $\sim$4.0 days after the second peak. The spectrum was obtained using the Double Spectrograph (DBSP, \citealt{OKE82}) mounted on the Palomar 200-inch Hale Telescope (P200). Both iPTF13beo spectra are shown in Figure \ref{spec_general} and are publicly available via the Weizmann Interactive Supernova data REPository (WISeREP\footnote{http://www.weizmann.ac.il/astrophysics/wiserep/}, \citealt{YAR12}).

The early spectrum is blue and is characterized by an absence of narrow or intermediate-width strong features. The redshift $z\approx0.091$ was estimated from a weak narrow unresolved H$\alpha$ emission, which seems to originate from the SN host galaxy. A line at $\lambda\approx4675$\,\AA{} coincides with the Half-Power Point (HPP) between the red and the blue SNIFS channels (\citealt{BUT13}) redshifted to the iPTF13beo host redshift and shown in Figure \ref{spec_general}. Therefore this line is probably an artifact produced by noise at the edges of the SNIFS channels. 

The late spectrum is also blue and exhibits some strong narrow and intermediate-width (FWHM $\approx$ 2000--5000\,\AA{}) He\,I and Si\,II emission features, which are zoomed in Figure \ref{spec_lines}. The narrow H$\alpha$ emission is also visible in the late spectrum and is also shown in Figure \ref{spec_lines}. The fact that the H$\alpha$ line in our spectra is unresolved, while the He\,I and Si\,II features are resolved, suggests that the H$\alpha$ emission originates in the host galaxy.

We assume that the narrow and intermediate-width spectral features in the late spectrum emerge from the SN ejecta interaction with a dense He-rich CSM. These features, unseen in the early spectrum taken during the SN rising phase, appear only after the second peak, once the SN ejecta hit the CSM. In addition, the late SN spectrum is similar to the SN 2006jc spectrum at phase +21 day, as shown in Figure \ref{spec_general}. iPTF13beo is thus another example of a Type Ibn (SN 2006jc-like) event, i.e., of a massive star exploding in a dense He-rich CSM.    

In addition to the weak and unresolved H$\alpha$ emission, there is an unresolved and yet weaker [O\,III] emission observed in both spectra, which is probably also associated with the host galaxy. Note, that the [O\,III] primary component at 5007\,\AA{} is mixed with the He\,I 5016\,\AA{} line, while the [O\,III] secondary component at 4959\,\AA{} is not contaminated. A very weak H$\beta$ emission, originating probably in the host galaxy as well, is observed only in the late spectrum, which may be attributed to a much better light-collecting power of the P200 telescope relative to that one of the U88 telescope. 

\section{Interpretation of the first peak} 

Another class of SNe interacting with a dense CSM are Type IIn events, characterized by the presence of narrow H emission lines (\citealt{SCH90,FIL97,KIE12}). \cite{OFE10} suggested that the early light curves of some Type IIn SNe are powered by shock breakout in a dense CSM (see also \citealt{CHE11}). One particular case, PTF 09uj, showed some evidence for a shock breakout from a dense circumstellar wind. The fast rise of the PTF 09uj ultra-violet (UV) light curve was explained by the SN shock breakout through a dense ($n\approx10^{10}$ cm$^{-3}$) CSM. PTF 09uj was characterized by a high mass loss rate ($\sim$0.1\,$M_{\odot}$\,yr$^{-1}$) prior to the explosion. The dense CSM surrounding PTF 09uj was thus suggested to originate from the pre-explosion stellar wind or outburst. Evidence for such pre-explosion re-brightening, presumably associated with mass-loss events, was recently uncovered for several Type Ibn and IIn SNe (\citealt{FOL07,PAS07,MAU13,OFE13c,OFE13d,PRI13,SMI13,OFE14}) and possibly also Type Ic SNe (\citealt{COR13}). 

We compared the iPTF13beo data with those of PTF 09uj and used the \cite{OFE10} order-of-magnitude model to derive various physical parameters of iPTF13beo. In all the calculations we used He opacity, $\kappa=0.2$\,cm$^2$\,g$^{-1}$. The shock velocity ($v_s\sim1.4\times10^4$\,km\,s$^{-1}$) was estimated from the iPTF13beo luminosity and rise-time, following Equation (4) in \cite{OFE10}. The distance from the star to the shell, in which the breakout occurs, was estimated using Equation (5) in \cite{OFE10} and was found to be $r_s\sim4\times10^{14}$\,cm. The CSM density at this radius was estimated to be $n\sim2\times10^{11}$ cm$^{-3}$ using Equation (6) in \cite{OFE10}. Assuming a wind density profile of $\rho\propto r^{-2}$ and integrating over distance, we estimate the total mass in the CSM within the radius $r_s$ to be $M\sim0.1$\,M$_{\odot}$. Assuming a wind velocity $v_w\approx3000$\,km\,s$^{-1}$ based on the line profiles, we estimate the mass-loss duration as $t\sim 15$\,days, implying the pre-explosion mass-loss rate was $\dot{M}\sim2.4$\,M$_{\odot}$\,yr$^{-1}$, i.e., a scenario of a sudden violent eruption. The exact ejected mass estimate depends on the ejected mass density profile and the deviation from the spherical symmetry. Therefore, the calculation here should be regarded as an order of magnitude estimate.

The physical properties of iPTF13beo mentioned above are comparable to the corresponding properties of PTF 09uj, except for a much higher mass-loss rate that lasted a much shorter time interval. The ejection of such an amount of matter at such a short time is similar to SN 2006jc-like pre-SN outbursts (\citealt{NAK06,PAS07,FOL07}) rather than to a stellar wind. Note that in the case of an outburst, the wind density profile assumption ($\rho\propto r^{-2}$) does not necessarily hold. 

Given the possible high mass-loss rate of the progenitor prior to the SN explosion, we searched for precursor events (e.g., \citealt{FOL07,PAS07,COR13,MAU13,PRI13,OFE13c,OFE13d,OFE14}). The search was conducted by co-adding PTF images (see \citealt{OFE13c} for details). In the 100 days prior to the SN explosion we did not find any sign for a precursor brighter than a limiting magnitude of $R=22.7$ mag (corresponding to an absolute magnitude of -15.4 mag). We note that iPTF13beo is likely too far away for a clear detection of a precursor by PTF.

\cite{KLE13} proposed another explanation for rapidly evolving SN light curves. They suggested that some rapidly rising and declining SNe may be Type Ib/c events with typical masses and energies, but with a very small amount of radioactive material ejected. This model could, in principle, explain the first peak of iPTF13beo, while the second peak would be due to the interaction with the dense CSM. The time scale of the first peak in this scenario will be set by the photon diffusion time in the CSM, which will completely smear the first peak. Therefore, this scenario is not likely to explain the case of iPTF13beo. Note, however, that the timescale of the first peak is poorly constrained.

\section{Conclusions}

iPTF13beo was first detected on 2013 May 19.349, $\sim$3 hours after the estimated explosion time. The youngest Type Ibn event discovered to date, iPTF13beo provides a unique opportunity to study the early photometric and spectroscopic evolution of this rare SN subtype.

The iPTF13beo $R$-band light curve is characterized by a very rapid rise ($\sim$2 days), which is faster than the rise of SN 1994I (a fast Type Ic), reaching a peak $R$-band absolute magnitude of about $-18.5$\,mag. The iPTF13beo light curve exhibits a double-peak structure, never before observed among other Type Ibn events. The first peak of the light curve is probably explained by the SN breakout, while the second peak is explained by a regular SN radioactive decay modified by the ejecta interaction with a dense He-rich CSM, which is further supported by the spectral follow-up results.

The declining phase of the iPTF13beo light curve is fitted well by both SN 1994I (Type Ic) and SN 2006jc (Type Ibn) light curves. The similarity of Type Ibn and fast Type Ic light curves was already noted by \cite{MAT00}, who discovered the first Type Ibn event - SN 1999cq. Hence, we conclude that the iPTF13beo light curve is consistent with a typical Type Ibn and fast Type Ic light curve during the declining phase, but has a slightly faster rise than a fast Ic light curve. The similar decline may indicate a low-mass ejecta and that the interaction with CSM does not dominate the $R$-band flux. 

The early spectrum, obtained during the rising phase, is blue and is characterized by absence of narrow spectral features, except of a weak H$\alpha$ and a very weak [O\,III] emission, which is assumed to originate from the host galaxy. 

The late iPTF13beo spectrum is also blue. Some He\,I and Si\,II narrow and intermediate-width (FWHM $\approx$ 2000--5000\,km\,s$^{-1}$) emission features emerge in the late spectrum. These features are $>$10 times wider than the previously mentioned weak H$\alpha$ emission, which is also present in the late spectrum, presumably arising from the host galaxy. Generally, the late spectrum is fitted well by the SN 2006jc spectrum. 

We compared the first peak of iPTF13beo to that of PTF 09uj (Type IIn) and found that if the first peak is assumed to be due to shock breakout in an optically thick CSM, the mass loss of iPTF13beo lasted for about a couple of weeks prior to the explosion at a rate of $\sim$2.4\,M$_\odot$\,yr$^{-1}$. This rapid mass loss at such a high rate corresponds to a SN 2006jc-like pre-explosion outburst rather than to a stellar wind.

Hence, we conclude that the spectral follow-up data, as well as the photometric data presented here, support the model of a massive stripped-envelope star exploding in and interacting with a dense He-rich circumstellar environment. 

\footnotesize
\paragraph*{}
ACKNOWLEDGMENTS: 
This paper is based on observations obtained with the
Samuel Oschin Telescope as part of the Palomar Transient Factory
project, a scientific collaboration between the
California Institute of Technology,
Columbia University,
Las Cumbres Observatory,
the Lawrence Berkeley National Laboratory,
the National Energy Research Scientific Computing Center,
the University of Oxford, and the Weizmann Institute of Science.
Some of the data presented herein were obtained at the W. M. Keck
Observatory, which is operated as a scientific partnership among the
California Institute of Technology, the University of California,
and NASA; the Observatory was made possible by the generous
financial support of the W. M. Keck Foundation.  We are grateful for
excellent staff assistance at Palomar, Lick, and Keck Observatories.
E.O.O. is incumbent of
the Arye Dissentshik career development chair and
is grateful to support by
a grant from the Israeli Ministry of Science and
the I-CORE Program of the Planning
and Budgeting Committee and The Israel Science Foundation (grant No 1829/12).

Avishay Gal-Yam acknowledges support by the EU/FP7 via ERC grant no.
307260, ISF, BSF, Minerva and GIF grants, the ``Quantum Universe''
I-Core program funded by the ISF and the Israeli Planning and Budgeting Committee,
and the Kimmel award.

Observations obtained with the SuperNova Integral Field Spectrograph on
the University of Hawaii 2.2-m telescope as part of the Nearby Supernova
Factory II project, a scientific collaboration between the Centre de
Recherche Astronomique de Lyon, Institut de Physique Nucl'eaire de Lyon,
Laboratoire de Physique Nucl'eaire et des Hautes Energies, Lawrence
Berkeley National Laboratory, Yale University , University, University of
Bonn , Max Planck Institute for Astrophysics, Tsinghua Center for
Astrophysics, and Centre de Physique des Particules de Marseille.

Nicolas Chotard acknowledges support from the Lyon Institute of Origins under grant ANR-10-LABX-66.

We would like to express our gratitude the anonymous referee for the valuable comments, that helped to improve this paper.

\end{document}